\documentclass[12pt]{article}
\usepackage{graphicx}
\newcommand{\AM}{\ensuremath{\mathcal{P}}}
\newcommand{\PA}{\ensuremath{\mathcal{C}}}
\newcommand{\RG}{\ensuremath{\mathcal{R}}}

\begin{document}
\title{Size and area of square lattice polygons}
\author{Iwan Jensen\thanks{e-mail: I.Jensen@ms.unimelb.edu.au} \\
Department of Mathematics \& Statistics, \\
The University of Melbourne, Victoria 3010, Australia}
\date{\today}
\maketitle
\bibliographystyle{plain}
\begin{abstract}
We use the finite lattice method to calculate the radius of gyration,
the first and second area-weighted moments of self-avoiding polygons
on the square lattice. The series have been calculated for polygons up to
perimeter 82. Analysis of the series yields high accuracy
estimates confirming theoretical predictions for the value of the
size exponent, $\nu=3/4$, and certain universal amplitude combinations.
Furthermore, a detailed analysis of the asymptotic form of the
series coefficients provide the firmest evidence to date for the 
existence of a correction-to-scaling exponent, $\Delta = 3/2$.
\end{abstract}

\section{Introduction}

A self-avoiding polygon (SAP) can be defined as a walk on a lattice
which returns to the origin and has no other self-intersections.
The history and significance of this problem is nicely discussed in 
\cite{Hug}. Generally SAPs are considered distinct up to translations, 
so if there are $p_n$ SAPs of perimeter length $n$ there are $2np_n$ walks 
(the factor of two arising since the walk can go in two directions). In 
addition to enumerations by perimeter, one can also enumerate polygons
by the enclosed area (or number of unit cells), or both perimeter
and area. Of particular interest are the first few area-weighted moments
of the perimeter generating function. Also of great interest is the 
mean-square radius of gyration, which measure the typical size of a SAP. 

This paper builds on the work of Enting \cite{Ent} who used transfer matrix
techniques to enumerate square lattice polygons by perimeter to 38 steps.  
This enumeration was later extended by Enting and Guttmann to 46 steps 
\cite{EG85} and then to 56 steps \cite{GE88a}. This latter work also included 
calculations of moments of the caliper size distribution. Hiley and Sykes 
\cite{HS} obtained the number of square lattice polygons by both area and 
perimeter up to perimeter 18. Enting and Guttmann extended the calculation to 
perimeter 42 \cite{EG90}. The radius of gyration was calculated for SAPs up 
to 28 steps by Privman and Rudnick \cite{PR}, using a technique based on
direct counting of compact site animals on the dual lattice. Recently, 
Jensen and Guttmann devised an improved algorithm for the enumeration of 
SAPs and extended the calculation to 90 steps \cite{JG99}. The work reported 
here is based on generalisations of this improved algorithm. This has enabled 
us to extend the calculation of the radius of gyration and the first two 
area-weighted moments to 82 step SAPs. The generalisation of the transfer
matrix technique to area-weighted moments is similar to the one used
by Conway \cite{Conway} in his calculation of series for percolation 
problems and lattice animals. The generalisation to the radius of gyration
has to our knowledge no counterpart in the published literature, and
represents a major advance in the design of efficient counting algorithms.
Previous calculations of the radius of gyration were based on direct
counting algorithms. The transfer matrix algorithm used in this paper
is exponentially faster and thus enables us to significantly extend
the series (see \cite{JG99} for further details).

The size exponent, $\nu$, for self-avoiding polygons is believed to
be identical to that of self-avoiding walks. This has been argued
theoretically from the connection between the energy-energy and spin-spin
correlation functions of the $n$-vector model in the limit $n \rightarrow 0$, 
and SAPs and SAWs, respectively \cite{deGennes,desCloizeaux}. Alternatively
it has also been obtained from real space renormalization group arguments
\cite{Family}. The exponent describing the growth of the mean area
of polygons of perimeter $n$ is expected to be $2 \nu$ \cite{LSF}. 
Intuitively this is not surprising since it just means that
the average area of a polygon is proportional to the square of the
radius of gyration. So one is merely finding that for
this problem the typical area and typical length scale match one another
nicely. These expectations have been confirmed reasonably accurately
by numerical work \cite{GE88a,EG90,PR}.

The functions we consider in this paper are: (i) the polygon generating
function, $\AM (u)= \sum p_n u^n$; (ii) $k^{th}$ area-weighted moments 
of polygons of perimeter $n$, $\langle a^k \rangle _n$; 
and (iii) the mean-square radius of gyration of polygons of perimeter $n$, 
$\langle R^2 \rangle _n$. These quantities are expected to behave as

\begin{eqnarray}\label{eq:coefgrowth}
p_n & = & B\mu^n n^{\alpha-3}[1+o(1)], \nonumber \\
\langle a^k \rangle _n & = & E^{(k)}n^{2k\nu}[1+o(1)], \\
\langle R^2 \rangle _n & = & D n^{2\nu}[1+o(1)], \nonumber
\end{eqnarray}
\noindent
where $\mu = u_c^{-1}$ is the reciprocal of the critical point of the
generating function, and $\alpha =1/2$ and $\nu = 3/4$ are known
exactly \cite{Nienhuis}, though non-rigorously. It is also known \cite{CG}
that the amplitude combination $E^{(1)}/D$ is universal, and that

\begin{equation}\label{eq:BDampl}
BD = \frac{5}{32 \pi ^2}\sigma a_0,
\end{equation}
\noindent
where $a_0$ is the area per site and $\sigma$ is an integer such that
$p_n$ is non-zero only if $n$ is divisible by $\sigma$. For the 
square lattice $a_0=1$ and $\sigma=2$. These predictions have
been confirmed numerically \cite{CG,Lin98}.

In the next section we describe the generalisation of the finite 
lattice method required in order to calculate the radius of gyration
and area-weighted moments of self-avoiding polygons. The results of
the analysis of the series are presented in Section~\ref{sec:analysis}.

\section{Enumeration of self-avoiding polygons \label{sec:sapenum}}

The method used to enumerate self-avoiding polygons in this work is based 
on the method devised by Enting \cite{Ent} for enumerations by perimeter
and uses the enhancements of Jensen and Guttmann \cite{JG99}. 
In the following we first very briefly outline the original method and
then show how to generalize it in order to calculate area-weighted moments
and the radius of gyration. Details of the algorithm can be found in the
papers cited above.

The first terms in the series for the perimeter generating function can be 
calculated using transfer matrix techniques to count the number of polygons 
spanning (in both directions) rectangles $W+1$ edges wide and $L+1$ edges 
long. The transfer matrix technique  involves drawing a boundary through the 
rectangle intersecting a set of $W+2$ edges. For each configuration of 
occupied or empty edges along the boundary we maintain a (perimeter) 
generating function for partially completed polygons. Polygons in a given 
rectangle are enumerated by moving the boundary so as to add one site at 
a time. Due to the obvious symmetry of the lattice one need only consider 
rectangles with $L \geq W$. Any polygon spanning such a rectangle has a 
perimeter of length at least $2(W+L)$.  By adding the contributions from all 
rectangles of width $W \leq W_{\rm max}$ (where the choice of $W_{\rm max}$ 
depends on available computational resources) and length 
$W \leq L \leq 2W_{\rm max}-W+1$, with contributions from rectangles 
with $L>W$ counted twice, the number of polygons per vertex of an infinite 
lattice is obtained correctly up to perimeter $n_{\rm max} =4W_{\rm max}+2$.
The number of configurations required as $W_{\rm max}$ is increased grows
exponentially as $\lambda^{W_{\rm max}}$, where $\lambda \simeq 2$ for 
the improved algorithm \cite{JG99}. In addition to the dominant
exponential growth in memory requirements there is a prefactor,
which is proportional to the number of terms $n_{\rm max}$.

\subsection{Area-weighted moments \label{sec:momenum}}

Area-weighted moments can easily be calculated from the perimeter
and area generating function

\begin{equation}\label{eqn:perimarea}
\PA (u,v)  =  \sum_{n,m} c_{n,m} u^{n}v^m,
\end{equation}
\noindent
where $c_{n,m}$ is the number of polygons with perimeter $n$ and area $m$.
From this we get the area-weighted generating functions,
\begin{equation}\label{eqn:areaweighted}
\AM^{(k)}(u)  =  (v\frac{\partial}{\partial v})^k \PA (u,v)|_{v=1} = 
           \sum_{n} \sum_{m}   m^k c_{n,m} u^{n} = \sum_{n} p_n^{(k)}u^{n},
\end{equation}
\noindent
and we define the average moments of area for a polygon with perimeter $n$
\begin{equation}\label{eqn:areamoment}
\langle a^k \rangle _n  = p_n^{(k)}/p_n^{(0)}= \sum_{m}   m^k c_{n,m}/p_n.
\end{equation}
\noindent

In order to calculate the moments of area through this approach we need to 
calculate a full two-parameter generating function, which generally will 
require a lot of computer memory. If we are only interested  in the first 
few moments there is a much more efficient approach \cite{Conway}. We simply 
replace the variable $v$ by $1+z$ thus obtaining the function
\begin{equation}\label{eqn:newvar}
F(u,z)  =  \sum_{n,m} c_{n,m} u^{n}(1+z)^m =  
\sum_{n,m} \sum_{k=0}^m \left(\!\!\begin{array}{cc}m\\ k\end{array}\!\!\right)
 c_{n,m} u^{n} z^m.
\end{equation}
\noindent
Let, $F_i (u)$, be the coefficient of $z^i$ in $F(u,z)$. Then we see that
\begin{eqnarray*} 
F_0(u)& = &\sum_{n,m}  c_{n,m} u^{n}= \AM(u), \\
F_1(u)& = &\sum_{n,m} m c_{n,m} u^{n}= \AM^{(1)}(u),\\ 
F_2(u)& = &\sum_{n,m} m(m\!-\!1)/2 c_{n,m} u^{n}= 
           [\AM^{(2)}(u)-\AM^{(1)}(u)]/2,
\end{eqnarray*} 
\noindent
and so on. Thus if we are only interested in the first and second moments
of area we can truncate the series $F(u,z)$ at second order in $z$
and find the relevant moments as $\AM^{(1)}(u)=F_1(u)$ and
$\AM^{(2)}(u)=2F_2(u)+F_1(u)$. The growth in memory requirements is still 
dominated by the exponential growth in the number of configurations.
However, we have managed to turn the calculation of these moments from a 
problem with a prefactor cubic in $W_{\rm max}$ (the area is proportional to
$W^2_{\rm max}$) into a problem
with a prefactor linear in  $W_{\rm max}$.

\subsection{Radius of gyration \label{sec:rgenum}}

In the following we show how the definition of the radius of gyration 
can be expressed in a form suitable for a transfer matrix calculation.
Note that we define the radius of gyration  according to the 
{\em vertices} of the SAP and that the number of vertices equals the 
perimeter length. The radius of gyration of $n$ points at positions
${\bf r}_i$ is 

\begin{equation} \label{eq:rg}
n^2 R^2_n = \sum_{i>j} ({\bf r}_i-{\bf r}_j)^2 =
(n\!-\!1)\sum_i (x_i^2+y_i^2)-2\sum_{i>j}(x_ix_j+y_iy_j).
\end{equation}

This last expression is suitable for a transfer matrix calculation. 
As usual \cite{PR} we actually calculate the generating function,
$\RG^2_g(u) = \sum_n p_n\langle R^2 \rangle _n n^2 u^n$. In order to
do this we have to maintain five partial generating functions
for each possible boundary configuration $\sigma$, namely

\begin{itemize}
\item $P(u)$, the number of (partially completed) polygons according 
to perimeter.
\item $R^2(u)$, the sum over polygons of the squared components of the 
distance vectors.
\item $X(u)$, the sum of the $x$-component of the distance vectors.
\item $Y(u)$, the sum of the $y$-component of the distance vectors.
\item $XY(u)$, the sum of the `cross' product of the components of the  
distance vectors, e.g., $\sum_{i>j}(x_ix_j+y_iy_j)$.
\end{itemize}

As the boundary line is moved to a new position  each boundary configuration 
$\sigma$ might be generated from several configurations $\sigma'$ 
in the previous boundary position. The partial generation functions are
updated as follows

\begin{eqnarray}
P(u,\sigma) & = & \sum_{\sigma'} 
  u^{n(\sigma')} P(u,\sigma'), \nonumber \\
R^2(u,\sigma) & = & \sum_{\sigma'}  
  u^{n(\sigma')}[R^2(u,\sigma')+\delta (x^2+y^2)P(u,\sigma')],\nonumber \\ 
X(u,\sigma) & = & \sum_{\sigma'}   
  u^{n(\sigma')}[X(u,\sigma)+ \delta xP(u,\sigma')], \\ 
Y(u,\sigma) & = & \sum_{\sigma'}   
  u^{n(\sigma')}[Y(u,\sigma)+ \delta yP(u,\sigma')], \nonumber \\ 
XY(u,\sigma) & = & \sum_{\sigma'}  
  u^{n(\sigma')} [XY(u,\sigma')+\delta xX(u,\sigma')
                   +\delta yY(u,\sigma')] \nonumber 
\end{eqnarray}
\noindent
where $n(\sigma')$ is the number of occupied edges added to the 
polygon and $\delta=\min (n(\sigma'),1)$. 

\subsection{Further particulars}

Finally a few remarks of a more technical nature. The number of 
contributing configurations becomes very sparse in the total set of 
possible states along the boundary line and as is standard in such 
cases one uses a hash-addressing scheme. Since the integer coefficients 
occurring in the series expansions become very large, the calculation 
was performed using modular arithmetic. Up to 8 primes were needed to 
represent the coefficients correctly. Further details and references are 
given in \cite{JG99}. The series for the radius of gyration and 
area-moments were calculated for SAPs with perimeter length up to 82. 
The maximum memory required for any given width did not exceed 2Gb.
The calculations were performed on an 8 node Alpha Server 8400 with a
total of 8Gb memory. The total CPU time required was about three days
per prime. Obviously the calculation for each width and prime are
totally independent and several calculations were done simultaneously.

In Table~\ref{tab:series} we have listed the series for the radius of 
gyration and first and second area-weighted moments. The series for the 
radius of gyration of course agree with the terms up to length 28 computed 
previously \cite{PR}, while the terms up to length 40 for the first area 
moment agree with the series in \cite{EG90}. The number of polygons of 
length $\leq 56$ can be found in \cite{GE88a} while those up to length 
90 were reported in \cite{JG99}.

\section{Analysis of the series \label{sec:analysis}}

The series listed in Table~\ref{tab:series} have coefficients which 
grow exponentially, with sub-dominant term given by a critical exponent.
The generic  behaviour is $G(u) =\sum_n g_n u^n \sim (1-u/u_c)^{-\xi},$ 
and hence the coefficients of the generating function 
$g_n \sim \mu^n n^{\xi-1}$, where $\mu = 1/u_c$. To obtain the singularity 
structure of the generating functions we used the numerical method of 
differential approximants \cite{Guttmann89}. In particular, we used this 
method to estimate the critical exponents (we already have very accurate 
estimates for $u_c$ from \cite{JG99}). Since all odd terms in the series 
are zero and the first non-zero term is $g_4$ we actually analysed the 
function $F(u)=\sum_n g_{2n+4} u^n$. Combining the relationship given 
above between the coefficients in a series and the critical behaviour of 
the corresponding generating function with the expected behaviour 
(\ref{eq:coefgrowth}) of the mean-square radius of gyration and moments 
of area yields the following prediction for their generating functions:

\begin{eqnarray}\label{eq:genfunc}
\RG^2_g (u)& = &\sum_n p_{2n+4}\langle R^2 \rangle _{2n+4}n^2 u^n =
    \sum_{n} r_n u^n \sim R(u)(1-u\mu^2)^{-(\alpha+2\nu)}, \\
\AM^{(k)} (u)& = &\sum_n p_{2n+4}\langle a^k \rangle _{2n+4} u^n =
    \sum_{n} a^{(k)}_n u^n \sim a^{(k)}(u)(1-u\mu^2)^{2-(\alpha+2k\nu)}.
\end{eqnarray}
\noindent
Thus we expect these series to have a critical point,
$u_c=1/\mu^2=0.14368062927(1)$, known to a very high degree of
accuracy from the analysis in \cite{JG99}, and as stated 
previously the exponent $\alpha=1/2$, while it is expected that
$\nu=3/4$.

Estimates of the critical point and critical exponents were obtained by 
averaging values obtained from second order $[L/N;M;K]$ inhomogeneous 
differential approximants. For each order $L$ of the inhomogeneous polynomial 
we averaged over those approximants to the series which used at least the 
first 80\% - 90\% of the terms of the series. We used only approximants 
where the difference between $N$, $M$, and $K$  didn't exceed 2.
Some approximants were excluded from the averages because the estimates were 
obviously spurious. The error quoted for these estimates reflects the spread 
(basically one standard deviation) among the approximants. Note that these 
error bounds should {\em not} be viewed as a measure of the true error as 
they cannot include possible systematic sources of error. In 
Table~\ref{tab:critexp} we have listed the results of our analysis.
It is evident that the estimates for $u_c$ and the critical exponents
are in agreement with the expected behaviour. There are only some minor
discrepancies in the fourth digit between the conjectured exponents and
the estimates. This discrepancy is readily resolved by looking at the
evidence in figure~\ref{fig:rgscrpexp}, where we have plotted
the estimates for the critical point and exponent of $\RG^2_g$. 
Each point in these figures represent an estimate obtained from a
specific second order differential approximant with the various points
obtained by varying the order of the polynomials in the approximants.
It is clear that the estimates have not yet settled down to their
asymptotic values and that they do converge towards the expected values
as the number of terms used by the approximants is increased.

Now that the exact values of the exponents has been confirmed we turn our 
attention to the ``fine structure'' of the asymptotic form of the
coefficients. In particular we are interested in obtaining accurate
estimates for the amplitudes $B$, $D$ and $E^{(1)}$. We do this
by fitting the coefficients to the assumed form (\ref{eq:coefgrowth}).

The asymptotic form of the coefficients $p_n$ of the polygon generating 
function has been studied in detail previously \cite{CG96,JG99}.
As argued in \cite{CG96} there is no sign of non-analytic 
corrections-to-scaling exponents to the polygon generating function 
and one therefore finds that

\begin{equation}\label{eq:sapasymp}
p_n = \mu^n n^{-5/2} \sum_{k=0} a_k/n^k.
\end{equation}
\noindent
This form was confirmed with great accuracy in \cite{JG99}. 
Estimates for the leading amplitude $B=a_0$ can thus be obtained by
fitting $p_n$ to  the form given in equation~(\ref{eq:sapasymp}). In order to 
check the behaviour of such estimates we did the fitting using from 
2 to 10 terms in the expansion. The results for the leading amplitude 
are displayed in figure~\ref{fig:sapampl}. We notice that all fits appear to 
converge to the same value as $n \rightarrow \infty$, and that, as more and 
more correction terms are added to the fits the estimates exhibits
less curvature and that the slope become smaller (although the fits using 
10 terms are a little inconclusive).  This is very strong evidence that 
(\ref{eq:sapasymp})  indeed is the correct asymptotic form of $p_n$. We 
estimate that $B=0.5623012(1)$.

The asymptotic form of the coefficients $r_n$ in the generating function
for the radius of gyration has not been studied previously. When fitting 
to a form similar to equation~(\ref{eq:sapasymp}), assuming that here are 
only analytic corrections-to-scaling, we find that the amplitudes of higher
order terms are very large and that the leading amplitude converge rather
slowly. This indicates that this asymptotic form is incorrect. We find 
that the coefficients fit better if we assume a leading non-analytic 
correction-to-scaling exponent $\Delta=3/2$. This result confirms 
the prediction of Nienhuis \cite{Nienhuis}. Note, that since the polygon 
generating function exponent $2-\alpha = 3/2$ a correction-to-scaling 
exponent $\Delta=3/2$ is perfectly consistent with the asymptotic form
(\ref{eq:sapasymp}). Because  $2-\alpha+\Delta$ is an integer the 
non-analytic correction term becomes part of the analytic background
term \cite{CG96}. We thus propose the following asymptotic form: 

\begin{equation}\label{eq:rgsasymp}
r_n = \mu^n n [BD + \sum_{k=0} a_k/n^{k/2}].
\end{equation}
\noindent
Alternative we could fit to the form
\begin{equation}\label{eq:rgsratasymp}
r_n/p_n = n^{7/2} [D + n^{5/2}\sum_{k=0} a_k/n^{k/2}].
\end{equation}
\noindent
In figure~\ref{fig:rgsampl} we show the leading amplitudes resulting from 
such fits while using from 1 to 10 terms in these expansions. Also shown
in these figures (solid lines) are the predicted exact value of $BD$, given
in equation~{\ref{eq:BDampl}, and the prediction for $D$ using the estimate 
for $B$ obtained above. As can be seen the leading amplitudes clearly 
converge towards their expected values and from these plots we can conclude 
that the prediction for $BD$ has been confirmed to at least 6 digit accuracy.
Assuming that  equation~({\ref{eq:BDampl}) is exact and using the very 
accurate estimate for $B$ we find that $D=0.05630944(1)$.

Fitting the coefficients for the area-weighted moments to asymptotic
forms similar to equations (\ref{eq:rgsasymp}) and  (\ref{eq:rgsratasymp})
above (only the leading exponent was changed
accordingly) leads to the estimates $E^{(1)}=0.141520(1)$ and
$E^{(2)}=0.0212505(4)$. 

As stated above the analysis of the polygon generating function is 
fully consistent with the prediction $\Delta = 3/2$. However, all one can 
conclude from the analysis is that, {\em if} non-analytic 
correction-to-scaling terms are present, the exponents have to be 
``half-integer'', so that the correction terms become part of the 
analytic background. The detailed analysis of the asymptotic form of the
coefficients in the generating functions for the radius of gyration and 
area-weighted moments provide the firmest evidence to date for the 
{\em actual existence} of a leading non-analytic correction to scaling 
exponent $\Delta = 3/2$, thus confirming the theoretical predictions 
made by Nienhuis \cite{Nienhuis}.

\section{Conclusion}

We have presented an improved algorithm for the calculation of the radius 
of gyration and area-weighted moments of self-avoiding polygons on the square 
lattice. This algorithm has enabled us to calculate these series for polygons 
up to perimeter length 82. Our extended series enables us to give very
precise estimate of the critical exponents, which are consistent with the 
exact values $\alpha = 1/2$ and $\nu = 3/4$. We also obtain a very precise
estimate for the amplitude $B=0.5623012(1)$. Analysis of the coefficients
of the radius of gyration series yielded results fully compatible with the 
prediction $BD=5/16\pi^2$. This allows us to obtain the very accurate
estimate $D=0.05630944(1)$. From the first area-weighted moment we
obtained the estimate $E^{(1)}=0.141520(2)$, which allows us to give a 
much improved estimate for the universal amplitude ratio 
$E^{(1)}/D=2.51326(3)$. We also find firm evidence for the existence
of a non-analytic correction-to-scaling term with exponent $\Delta = 3/2$.

\section*{E-mail or WWW retrieval of series}

The series for the various generating functions studied in this paper
can be obtained via e-mail by sending a request to 
I.Jensen@ms.unimelb.edu.au or via the world wide web on the URL
http://www.ms.unimelb.edu.au/\~{ }iwan/ by following the instructions.

\section{Acknowledgements}

Thanks to Tony Guttmann for many valuable comments on the
manuscript and the series analysis. 
Financial support from the Australian Research Council is
gratefully acknowledged.

\eject

\begin{figure}
\begin{center}
\includegraphics{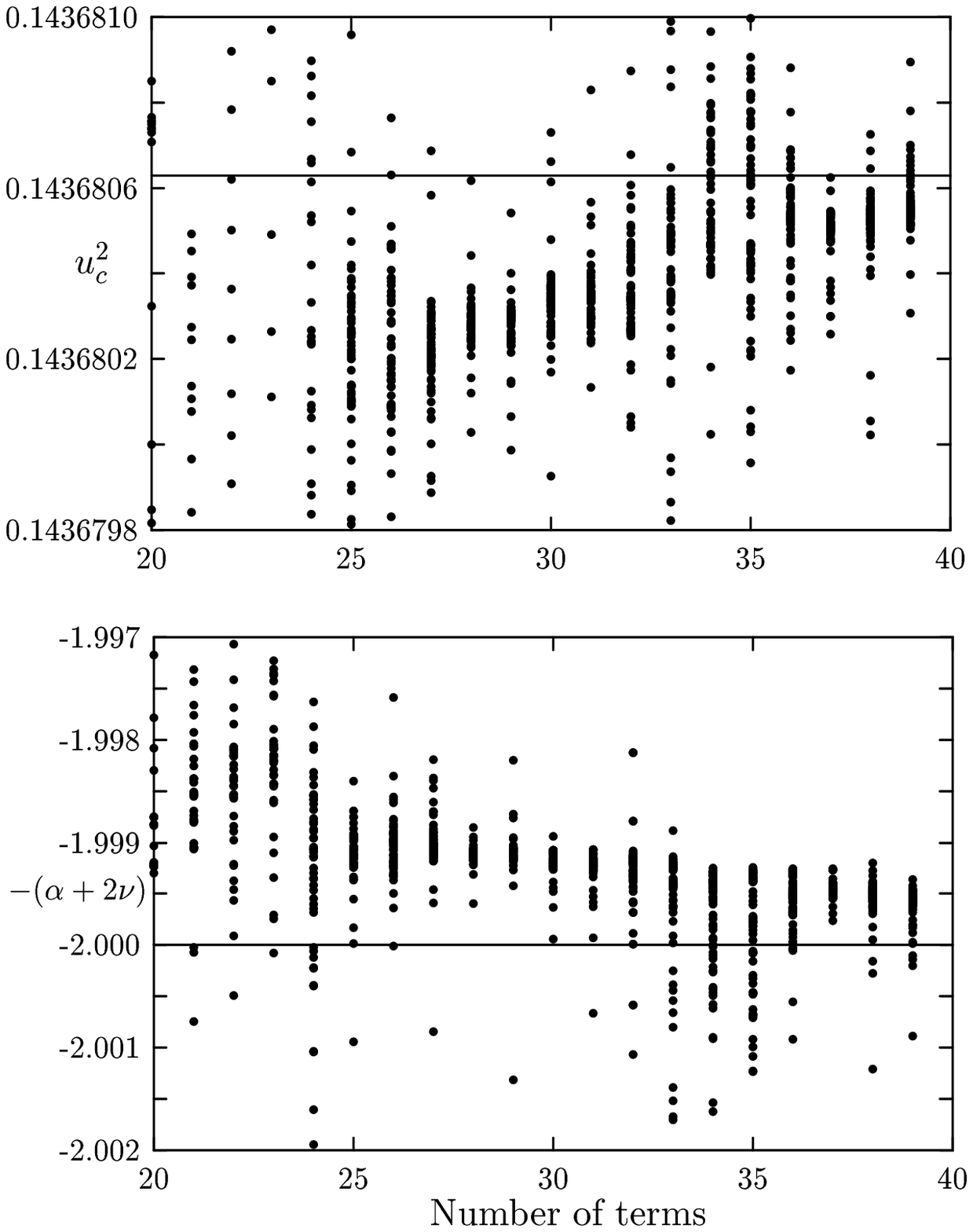}
\end{center}
\caption{\label{fig:rgscrpexp} Estimates for the critical point and
exponent of the generating function for the radius of gyration of square 
lattice polygons as a function of the number of terms used by the 
second order differential approximants. The solid lines indicate the expected
values $u_c=0.14368062927(1)$ and $\xi =-(\alpha+2\nu)=-2$.}
\end{figure}

\begin{figure}
\begin{center}
\includegraphics{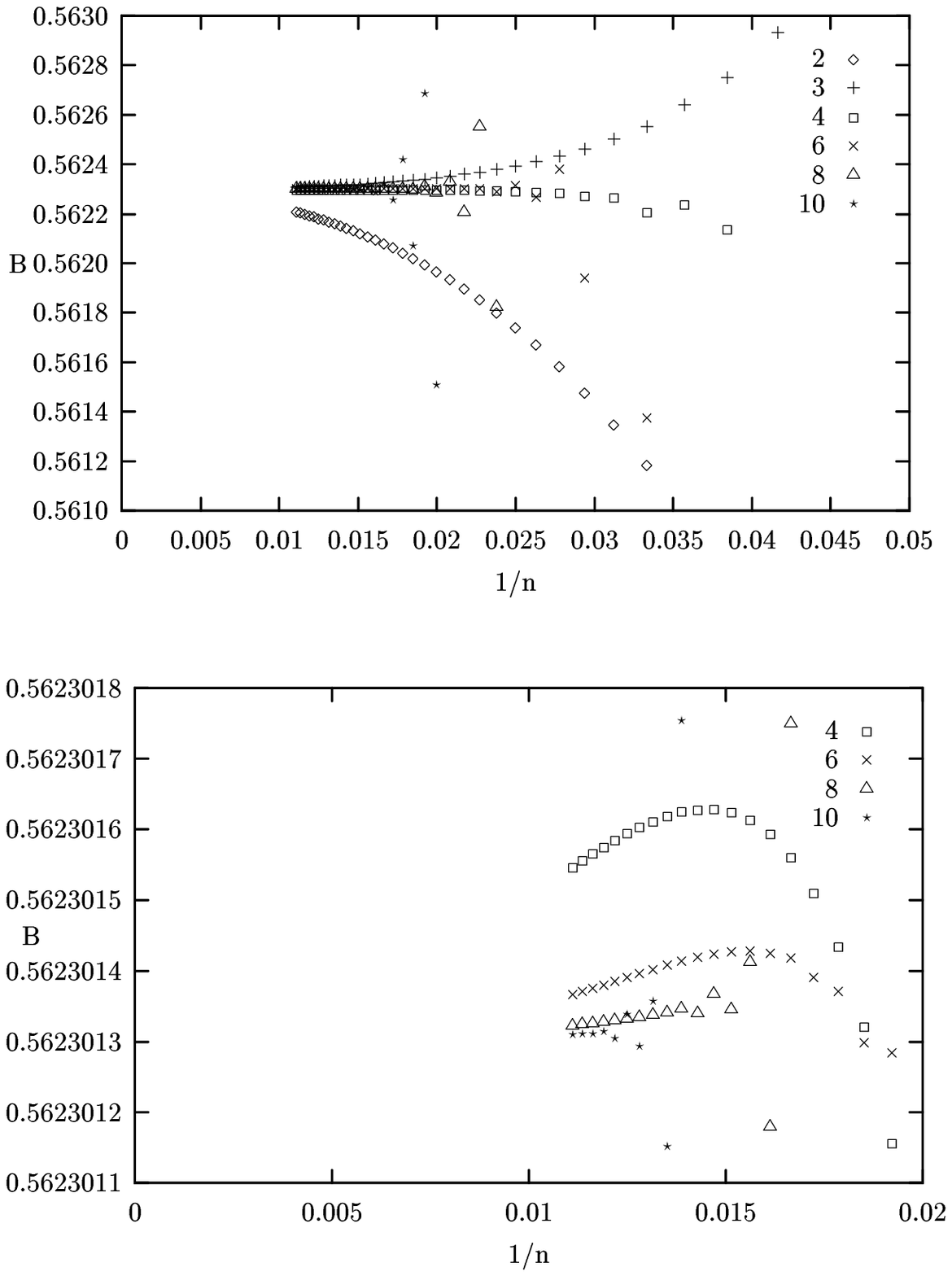}
\end{center}
\caption{\label{fig:sapampl} Estimates for the leading amplitude $a_0=B$
of square lattice polygons as a function of $1/n$. Each data set is
obtained by fitting $p_n$ to the form given in 
equation~(\protect \ref{eq:sapasymp}) using from 2 to 10 correction terms.
The lower panel displays a detailed look at the data in the upper panel.}
\end{figure}

\begin{figure}
\begin{center}
\includegraphics[scale=0.8]{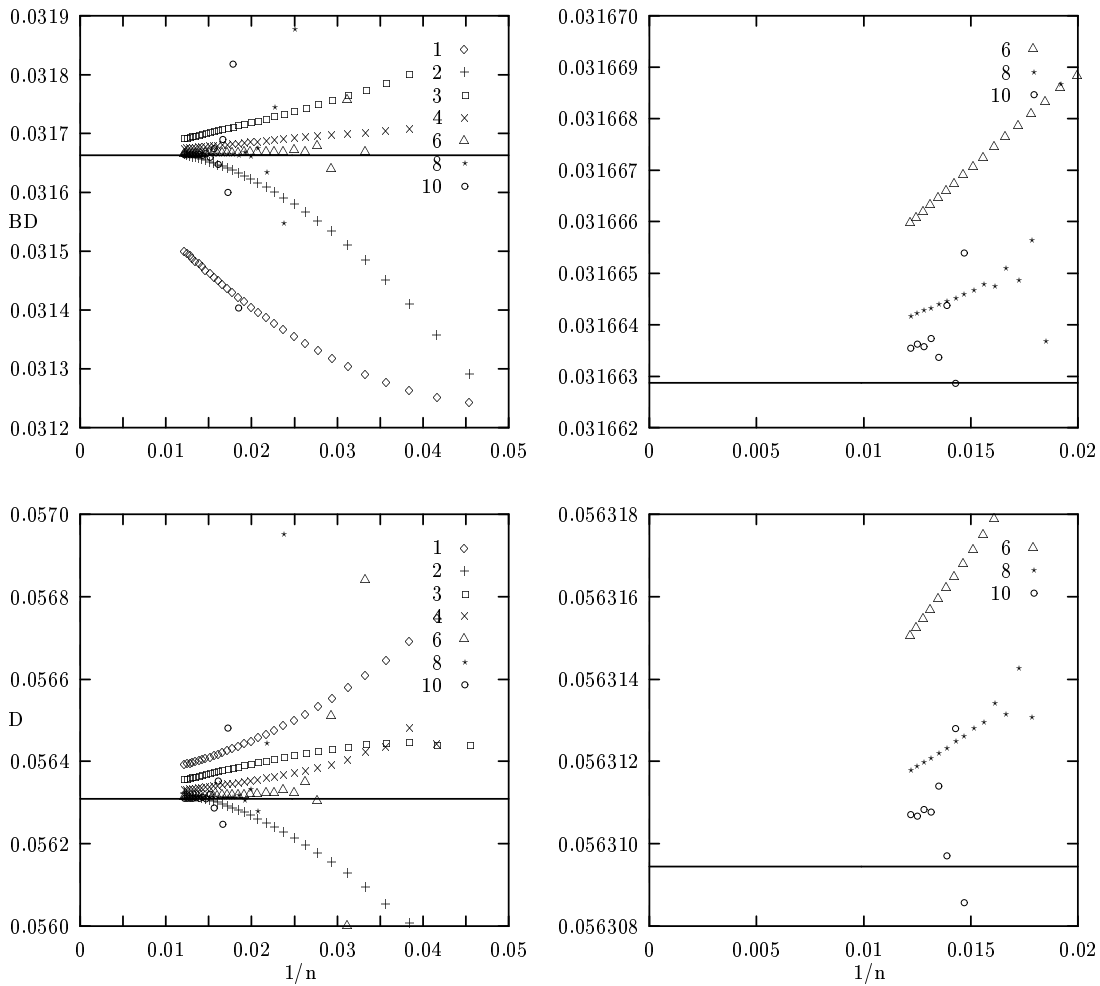}
\end{center}
\caption{\label{fig:rgsampl} Estimates for the leading amplitude $BD$
and  $D$ of the radius of gyration of square lattice polygons as a 
function of $1/n$. Each data set in the top panels is obtained by fitting 
the coefficients $r_n$ of the radius of gyration generating function to 
the form given in equation~(\protect \ref{eq:rgsasymp}), using from 1 
to 10 correction terms. Each data set in the bottom panels is obtained by 
fitting $r_n/p_n$ to the form given in 
equation~(\protect \ref{eq:rgsratasymp}).
The right panels are a detailed look at the data in the left panels.
The solid lines indicate the expected values given in the text.}
\end{figure}

\eject

\begin{table}
\caption{\label{tab:series} The mean-square radius of gyration, first and 
second area-moments of $n$-step self-avoiding polygons on the square lattice. 
Only non-zero terms are listed.}
\begin{center}
\tiny
\begin{tabular}{rrrr} \hline \hline
$n$ & $p_n n^2\langle R^2 \rangle _n$ & 
$p_n\langle a \rangle _n$ & $p_n\langle a^2 \rangle _n$ \\ \hline 
 4   &  8  & 1 & 1  \\
 6   &  66 & 4 & 8  \\
 8   &  600  & 22 & 70  \\
10   &  5164 & 124 & 560  \\
12   &  42872 & 726 & 4358  \\
14   &  346828 & 4352 & 33160  \\
16   &  2754056  & 26614 & 248998  \\
18   &  21549780  & 165204 & 1851040  \\
20   &  166626744  & 1037672 & 13655432  \\
22   &  1275865332  &  6580424 & 100126648  \\
24   &  9690096824  &  42062040  & 730548788  \\
26   &  73090383120  & 270661328  & 5308524968   \\
28   &  548064459968  &  1751614248  & 38442000664  \\
30   &  4088719617824  &  11391756176  & 277565593032  \\
32   &  30367415294800  &  74406502814  &  1999068564026 \\
34   &  224659143155964  &  487838450116  & 14365917755936  \\
36   &  1656259765448200  &  3209229661682  & 103038218758426  \\
38   &  12172580326973688  &
  21175301453040  & 737765745264544  \\
40   &  89212147340159520  &
  140097533633112  & 5274413814993896  \\
42   &  652183776123444404  &
  929160187609096  & 37655943519835560  \\
44   &  4756877451862073312  &
  6176075676719784  &  268506373782824280 \\
46   &  34623252929242595840  &
  41135052992574928  & 1912438211281990104  \\
48   &  251526960780642980968  &
 274482801972069490  & 13607405560541031042 \\
50   &  1824061566724351292496  &
 1834665820375683428  & 96728883661202188552 \\
52   &  13206639904144205117592  &
  12282315178525359966 & 687010148492686667614 \\
54   &  95476389002729304216548  &
 82344395405972692656  & 4875571799890192459056 \\
56   &  689283065294740945143208  &
  552806313387704627982  & 34575571741149137524846 \\
58   &  4969805963839723557919424  &
 3715834986939390916244  &245029144855912573003776  \\
60   &  35789811145967164348552960  &
 25006203000374020526746  & 1735367234605432029439794  \\
62   &  257449325423816274956954508  &
 168466668960946012707912  & 12283126555855361655011856 \\
64   &  1849981836861769186990365288  &
 1136122707072612282498874  & 86893466632100569644163186 \\
66   &  13280506839637150191613774736  &
 7669275741518968346891172  & 614385797629196735502076968 \\
68   &  95248670945282200958664147712  &
 51817515409677258092083006  & 4341950222145487318409546446 \\
70   &  682533032784692897614712920788  &
 350404221555935013278573224  &  30671194434233707728683946784 \\
72   &  4886864684580008620898035643960  &
 2371438542131929578320200646  & 216565948566766116053230547838\\
74   &  34962179240623880562564354461036  &
 16061466455829089444235194204  & 1528529336761773075102657075616 \\
76   &  249946063483045736235271147799248  &
  108860864860439323866007261128 & 10784279532727353410458586600848 \\
78   &  1785625611982607482936563853493112  &
 738338427155234332385671368928 & 76059086282576056911156299311952 \\
80   &  12748122227351375676612377672210416  &
 5010964557143508508512736679936  &  536243262589039476652829061618528 \\
82   &  90955298658999234326739061737970500  &
 34029495976431151261075225822320  &  3779470144925357385934811283997288 \\
\hline\hline
\end{tabular}
\end{center}
\end{table}

\begin{table}
\caption{\label{tab:critexp} Estimates for the critical point
$u_c$ and exponents obtained from second order differential approximants to 
the series for the radius of gyration, first and second moments of area
of square lattice self-avoiding polygons. 
$L$ is the order of the inhomogeneous polynomial.}
\scriptsize
\begin{center}
\begin{tabular}{lllllll} \hline \hline
Series:    &  \multicolumn{2}{c}{$\RG^2_g (u)$} & 
       \multicolumn{2}{c}{$\AM^{(1)} (u)$} & 
       \multicolumn{2}{c}{$\AM^{(2)} (u)$} \\ \hline 
$L$ &\multicolumn{1}{c}{$u_c$} & \multicolumn{1}{c}{$-(\alpha+2\nu)$} & 
\multicolumn{1}{c}{$u_c$} & \multicolumn{1}{c}{$2-(\alpha+2\nu)$}  &
\multicolumn{1}{c}{$u_c$} & \multicolumn{1}{c}{$2-(\alpha+4\nu)$} \\ \hline 
 0 & 0.14368045(10)  & -1.99941(23)
   & 0.143680543(86)& 0.00027(19)
   & 0.143680502(35)& -1.499523(78) \\
 1 & 0.14368057(14)& -1.99976(54)
   & 0.143680556(33)& 0.000279(78)
   & 0.143680539(19)& -1.499662(62) \\
 2 & 0.14368063(13)& -1.99997(58)
   & 0.143680558(31)& 0.000266(66)
   & 0.143680535(22)& -1.499658(86) \\
 3 & 0.14368048(11)& -1.99948(21)
   & 0.143680562(25)& 0.000267(48)
   & 0.143680530(20)& -1.49958(21) \\
 4 & 0.143680540(71)& -1.99956(18)
   & 0.143680567(25)& 0.000253(57)
   & 0.143680541(21)& -1.499636(57) \\
 5 & 0.143680553(60)& -1.99959(18)
   & 0.143680566(27)& 0.000254(66)
   & 0.143680545(15)& -1.499654(28) \\
 6 & 0.143680542(33)& -1.999544(92)
   & 0.143680577(13)& 0.00018(17)
   & 0.143680542(15)& -1.499649(58) \\
 7 & 0.14368046(10)& -1.99942(14)
   & 0.143680564(23)& 0.000255(58)
   & 0.143680542(16)& -1.499658(64) \\
 8 & 0.143680511(43)& -1.999474(96)
   & 0.143680568(21)& 0.000246(48)
   & 0.143680539(17)& -1.499650(57) \\
 9 & 0.143680527(64)& -1.99952(15)
   & 0.1436805828(92)& 0.000214(30)
   & 0.143680548(28)& -1.499674(71) \\
10 & 0.143680511(41)& -1.999472(94)
   & 0.143680572(17)& 0.000238(43)
   & 0.143680544(16)& -1.499665(52) \\
\hline \hline
\end{tabular}
\end{center}
\end{table}

\end{document}